\documentclass[a4paper,11pt,twoside]{article}

\usepackage{geometry} 
\usepackage{amsmath}
\usepackage{amssymb}
\usepackage{color}
\usepackage{graphicx}

\newcommand{\GeV}{\,\mathrm{GeV}}

\newcommand{\fracwithdelims}[4]{\left#1 \frac{#3}{#4} \right#2}
\newcommand{\ord}[1]{\mathcal{O}\left( #1 \right)}

\newcommand{\vev}[1]{\left\langle #1\right\rangle}
\newcommand{\Fig}[1]{Fig.~\ref{fig:#1}}

\newcommand{\Eq}[1]{Eq.~(\ref{eq:#1})}
\newcommand{\eq}[1]{eq.~(\ref{eq:#1})}
\newcommand{\Eqs}[1]{Eqs.~(\ref{eq:#1})}
\newcommand{\eqs}[1]{eqs.~(\ref{eq:#1})}

\newcommand{\omi}[1]{}

\newlength{\myem}
\newcommand{\sep}[1]{#1}
\newcounter{mysubequation}[equation]
\renewcommand{\themysubequation}{\alph{mysubequation}}
\newcommand{\mytag}{\stepcounter{mysubequation}%
\tag{\theequation\protect\sep{\themysubequation}}}
\newcommand{\globallabel}[1]{\refstepcounter{equation}\label{#1}}

\makeatletter
\renewcommand{\section}{\@startsection{section}{1}{0em}%
        {-3.5ex \@plus -1ex \@minus -.2ex}% 
        {2.3ex \@plus.2ex}%
        {\normalfont\large\bfseries}}
\renewcommand{\subsection}{\@startsection{subsection}{2}{0em}%
        {-3.25ex\@plus -1ex \@minus -.2ex}%
        {1.5ex \@plus .2ex}%
        {\normalfont\bfseries}}
\renewcommand{\subsubsection}%
        {\@startsection{subsubsection}{3}{0em}%
        {-3.25ex\@plus -1ex \@minus -.2ex}%
        {1.5ex \@plus .2ex}%
        {\normalfont\itshape}}
\makeatother

\newcommand{\SISSA}{SISSA/ISAS and INFN, I--34013 Trieste, Italy}

\newcommand{\preprintnumber}{%
SISSA--49/2006/EP}
 
\newcommand{\titletext}{Flavour from accidental symmetries} 
\newcommand{\authortext}{\large Luca Ferretti$^{\, a}$, Stephen F. King$^{\, b}$ and Andrea Romanino$^{\, a}$
\medskip\\\em\normalsize 
$\mbox{}^a$ \SISSA
\\[0.1\baselineskip] 
$\mbox{}^b$ School of Physics and Astronomy, University of Southampton, \\
        Southampton, SO17 1BJ, U.K.}
\newcommand{\abstracttext}{We consider a new approach to fermion masses and mixings in which no special ``horizontal'' dynamics is invoked to account for the hierarchical pattern of charged fermion masses and for the peculiar features of neutrino masses. The hierarchy follows from the vertical, family-independent structure of the model, in particular from the breaking pattern of the Pati-Salam group. The lightness of the first two fermion families can be related to two family symmetries emerging in this context as accidental symmetries.}

\title{
\normalsize
\begin{tabular}[t]{l}%\hepnumber\\
\end{tabular}
\hspace*{\fill}
\begin{tabular}[t]{l}\preprintnumber\end{tabular}
\vspace{3\baselineskip}\\\Large\bfseries\titletext\bigskip}
\author{\begin{minipage}[t]{0.8\textwidth}
\normalsize\centering\authortext
\end{minipage}}
\date{}

\begin{document}

\bigskip
\maketitle
\begin{abstract}\normalsize\noindent
\abstracttext
\end{abstract}\normalsize\vspace{\baselineskip}

\section{Introduction}

The origin of the peculiar pattern of fermion masses and mixing might appear more or less transparent at low scale depending on the degree of understanding of the full theory it requires. 
%For example, the pattern we observe might be inextricably related to the ultimate understanding of fundamental interactions, which would force the poor phenomenologist to patiently wait for that understanding to be served. 
Most approaches to the problem rely on the possibility that a full understanding is not required and the pattern of fermion masses and mixings follows from a ``factorizable'' dynamical principle associated to the ``horizontal'' family indices. In this paper we discuss the possibility that not even such a dynamics needs to be known, or exists at all, and the peculiar fermion mass pattern we observe simply follows 
from the fact that one heavy vectorlike family of fields turns out to be lighter than the rest of the heavy fields. The couplings of this lighter heavy family with the light families will not be constrained by any symmetry or alternative mechanism imposed on the theory. They will instead all be of order one, perhaps determined by some fundamental theory we do not need to know, and the charged fermion hierarchy will follow from the hierarchy in the breaking of the vertical gauge structure of the theory, in particular from the breaking of the Pati-Salam (PS) gauge group~\cite{Pati:73a}. Chiral symmetries acting on family indices protecting the masses of the first two fermion families emerge in this context as accidental symmetries. 

In Section~\ref{sec:motivations} we motivate the structure of the model and in particular the choice of the left-right (LR) symmetric and Pati-Salam (PS) gauge groups. In Section~\ref{sec:analysis}, which is supposed to be self-contained, we define in detail the model and sistematically analyze it. Supersymmetry is assumed throughout the paper. 

\section{A bottom-up approach to flavour from accidental symmetries}
\label{sec:motivations}

\subsection{Messenger dominance}

Let $\psi_i=q_i, u^c_i,d^c_i,l_i,n^c_i,e^c_i$, $i=1,2,3$ denote the three light SM families in Weyl notations, including three singlet neutrinos, and let $h=h_u,h_d$ denote the light Higgs. As usual, the lightness of the three SM families (except possibly the singlet neutrinos) is guaranteed by their chirality with respect to the SM group, while additional degrees of freedom are allowed to be much heavier because they come in vectorlike representations of the SM group. As anticipated in the introduction, the pattern of fermion masses arises in our model from the existence of a single relatively light vectorlike family of ``messengers'' $\Psi+\overline\Psi$, with $\Psi=Q, U^c,D^c,L,N^c,E^c$, and from the breaking pattern of the gauge group. We also consider the possibility of heavy Higgs messenger fields $H=H_u,H_d$. 

Since $\Psi$ has the same SM quantum numbers as $\psi_i$, we use a discrete $\mathbf{Z}_2$ symmetry to tell the light families from the heavy one. The light fields $\psi_i$, $h$ are $\mathbf{Z}_2$-odd, while the messengers are even. In the unbroken limit, the light families are massless, while the messengers fields $\Psi,\overline\Psi,H$ are allowed to be superheavy\footnote{The SM Higgs $h$ is of course in principle also allowed to be heavy. We do not address this $\mu$-problem here. }. Yukawa couplings for the light fields are forbidden by the $\mathbf{Z}_2$ symmetry. In order to break it, we then also include a SM-singlet $\mathbf{Z}_2$-odd chiral field $\phi$. Its scalar component will get a vacuum expectation value (vev) at a heavy scale not far from the messenger scale. Needless to say, the $\mathbf{Z}_2$ symmetry is not a family symmetry, as it does not tell the three families apart, all being odd under it. This is similar to what done in~\cite{Barr:81a}, where the hierarchical pattern of fermion masses was also addressed without the use of family symmetries. 

\begin{figure}
\begin{center}
\includegraphics[width=\textwidth]{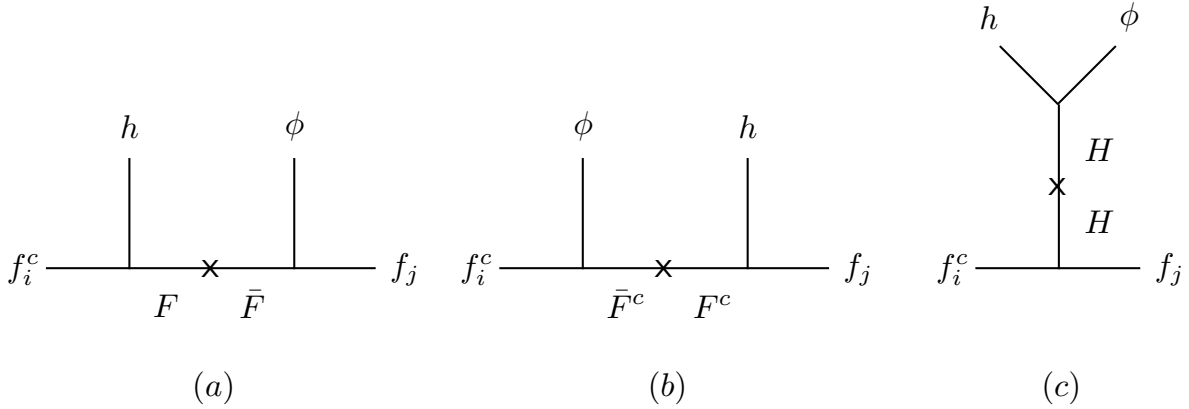}
\end{center}
\caption{Messenger exchanges contributing to the operator $(\phi/M)\psi_i\psi_jh$ in the effective theory below the messenger scale. $F$, $f$ refer to electroweak doublets, while $F^c$, $f^c$ refer to electroweak singlets.}
\label{fig:diagrams}
\end{figure}

Once $\phi$ gets a vev, the light and heavy fermions mix, which gives rise to the SM Yukawa couplings. In the limit in which the vev is smaller than the mass of the heavy messengers, $\vev{\phi}\ll M$, the Yukawa couplings of the light fermions can be seen to arise from higher dimensional operators in the effective theory below the scale $M$. This limit does not always hold in our model, as we will see, but it is useful for illustrative purposes and will be used in this Section. The exact treatment is postponed to Section~\ref{sec:analysis}. At the lowest order, the relevant operators are in the form $(\phi/M)\psi_i\psi_jh$ and they arise from the three diagrams in  \Fig{diagrams}. 

If the three contributions in \Fig{diagrams} are comparable and if the couplings involved are uncorrelated, we expect the fermion masses of the three families to be comparable. On the other hand, in the limit in which one of the three exchanges dominates (because the corresponding messenger is lighter) one family turns out to be heavier and a hierarchy is generated. This mechanism has several interesting features. The ``horizontal'' hierarchy among different families follows from a ``vertical'' hierarchy among messengers belonging to the same family, as in~\cite{Barr:81a}. As a consequence, the interfamily hierarchy can be attributed to the breaking pattern of the gauge group. Moreover, we will see that a two step breaking of the gauge group below the cutoff of the theory is sufficient to account for the complex hierarchical structure of charged fermions. We will also see that in spite of the absence of small coefficients, the CKM mixing angles will turn out to be small, while in the neutrino sector an attractive mechanism is available to give rise to a naturally large atmospheric mixing between normal hierarchical neutrinos. 

Let us see how this works in greater detail. Let us concentrate on the two heavier families and let us also neglect for the time being the Higgs exchanges in \Fig{diagrams}. We will discuss their role in connection to the first family masses in Section~\ref{sec:analysis}. In compact notations, the most general renormalizable superpotential is (we illegally use the same notation for the chiral superfield and its ``$R_P$-even'' component)
\begin{equation}
\label{eq:lagrangian}
W = M\bar\Psi\Psi +\alpha_i\bar\Psi\psi_i\phi +\lambda_i\Psi\psi_ih, 
\end{equation}
where
\begin{gather}
\label{eq:compact}
 M\bar\Psi\Psi \equiv M_Q \bar Q Q + M_U\bar U^c U^c +M_D\bar D^c D^c + M_L \bar L L + M_N\bar N^c N^c +M_E\bar E^c E^c  \notag \\
 \alpha_i\bar\Psi\psi_i\phi \equiv \alpha^Q_i\bar Q q_i\phi +\alpha^U_i\bar U^c u^c_i\phi +\alpha^D_i\bar D^c d^c_i\phi+  \alpha^L_i\bar L l_i\phi +\alpha^N_i\bar N^c n^c_i\phi +\alpha^E_i\bar E^c e^c_i\phi \\
\begin{split}
\lambda_i\Psi\psi_i h \equiv \lambda^{Qu}_i Qu^c_ih_u +\lambda^{Uq}_i U^c q_i h_u +\lambda^{Qd}_i Qd^c_ih_d +\lambda^{Dq}_i D^c q_i h_d + \\ 
\lambda^{Ln}_i Ln^c_ih_u +\lambda^{Nl}_i N^c l_i h_u +\lambda^{Le}_i Le^c_ih_d +\lambda^{El}_i E^c l_i h_d
\end{split} \; . \notag
 \end{gather}
%\begin{gather}
%\label{eq:compact}
% M\bar\Psi\Psi \equiv M_Q \bar Q Q + M_U\bar U^c U^c +M_D\bar D^c D^c +\ldots  \notag \\
% \alpha_i\bar\Psi\psi_i\phi \equiv \alpha^Q_i\bar Q q_i\phi +\alpha^U_i\bar U^c u^c_i\phi +\alpha^D_i\bar D^c d^c_i\phi+\ldots \\
% \lambda_i\Psi\psi_i h \equiv \lambda^{Qu}_i Qu^c_ih_u +\lambda^{Uq}_i U^c q_i h_u +\lambda^{Qd}_i Qd^c_ih_d +\lambda^{Dq}_i D^c q_i h_d +\ldots \; . \notag
% \end{gather}
No family symmetry or other dynamical constraint is imposed on the couplings. As a consequence, the dimensionless parameters in eq.~(\ref{eq:compact}) are all assumed to be $\ord{1}$ and uncorrelated.  
 When $\phi$ gets a vev,  the heavy and light fermions mix, which gives rise to the quark Yukawa matrices $Y^U$ and $Y^D$. In the limit $\vev{\phi}\ll M$ (and in the RL convention for the Yukawas)
\globallabel{eq:yukawas}
\begin{align}
-Y^U_{ij} &= \lambda^{Qu}_i\alpha^Q_j\frac{\vev{\phi}}{M_Q} 
+\alpha^U_i\lambda^{Uq}_j\frac{\vev{\phi}}{M_U} \mytag \\
-Y^D_{ij} &= \lambda^{Qd}_i\alpha^Q_j\frac{\vev{\phi}}{M_Q} 
+\alpha^D_i\lambda^{Dq}_j\frac{\vev{\phi}}{M_D} . \mytag 
\end{align}
Let us first consider the matrix $Y^U$. The up quark is massless, since $Y^U$ has rank two. If $M_Q\sim M_U$, the charm mass is expected to be of the same order of the top quark mass. This is because no horizontal hierarchy nor alignment is forced among the family dependent parameters $\alpha^Q_i$, $\alpha^U_i$, $\lambda^{Qu}_i$, $\lambda^{Uq}_i$. However, in the limit in which one of the terms in eq.~(\ref{eq:yukawas}a) dominates, the charm mass gets suppressed, as one messenger cannot give a mass to more than one family. A small $V_{cb}$ angle is only guaranteed if the $Q$ exchange is dominant in both the up and down quark sectors\footnote{This is true unless appropriate correlations are forced between the $U$ and $D$ coefficients, see below.}. We refer to this hypothesis as ``left-handed dominance''. We have then generated an inter-family hierarchy in terms of order parameters associated to the intra-family messenger structure, $M_Q/M_U, M_Q/M_D\ll 1$. The mechanism at work behind the explicit discussion above has to do with accidental flavour symmetries emerging in specific limits. First of all the discussion above holds in the limit in which the first family is massless. Such a limit, which will be defined in Section~\ref{sec:analysis}, implies the presence of an accidental chiral symmetry protecting the first family. Moreover, a second accidental symmetry protecting the masses of the of the second family fermions emerges in the limit in which $M_U$, $M_D$ become heavy.

A closer look to the textures obtained shows that in this framework the features of the fermion masses and mixings are best interpreted in the context of a Pati-Salam extension of the standard model group, as we now see. 

\subsection{$V_{us}$ and SU(2)$_R$ symmetry}

In order to write the Yukawa matrices in a simple form, we note that it is possible to choose a basis in the $q_i$, $u^c_i$, $d^c_i$ flavour space such that $\alpha^Q_{1,2}=\lambda^{Qu}_{1,2} = \lambda^{Qd}_{1,2} = 0$. We can then also rotate the ``1,2'' fields to set $\alpha^U_1 = \alpha^D_1 =\lambda^{Uq}_1 = 0$. If the dimensionless coefficients were of the same order and uncorrelated in the initial basis, we expect the non-vanishing coefficient to be still of the same order and uncorrelated in the new basis. The quark Yukawa matrices can now be written as
\globallabel{eq:y1}
\begin{gather}
Y^U = \begin{pmatrix}
0 & 0 & 0 \\
0 & r^U_2a^U_2\epsilon_U & r^U_3a^U_2\epsilon_U \\
0 & r^U_2a^U_3\epsilon_U & 1
\end{pmatrix}
\alpha^Q_3\lambda^{Qu}_3\frac{\vev{\phi}}{M_Q}, \mytag \\
Y^D = \begin{pmatrix}
0 & 0 & 0 \\
r^D_1a^D_2\epsilon_D & r^D_2a^D_2\epsilon_D & r^D_3a^D_2\epsilon_D \\
r^D_1a^D_3\epsilon_D & r^D_2a^D_3\epsilon_D & 1
\end{pmatrix}
\alpha^Q_3\lambda^{Qd}_3\frac{\vev{\phi}}{M_Q}, \mytag
\end{gather}
where $\epsilon_U=M_Q/M_U$, $\epsilon_D=M_Q/M_D\ll 1$ , while $r^U_i = \lambda^{Uq}_i/\lambda^{Qu}_3$, $r^D_i = \lambda^{Dq}_i/\lambda^{Qd}_3$, $a^U_i=\alpha^U_i/\alpha^Q_3$, $a^D_i=\alpha^D_i/\alpha^Q_3\sim \ord{1}$ or vanishing. 

A few remarks are in order. First of all, we note that \eqs{y1} give
\begin{equation}
\label{eq:prediction1}
\frac{m_s}{m_b}\approx r^D_2a^D_2\epsilon_D \sim r^D_2a^D_3\epsilon_D \approx |V_{cb}| ,
\end{equation}
in agreement with data. In contrast, flavour symmetries often give $m_s/m_b\sim |V_{cb}|^2$, unless non-abelian symmetries~\cite{Barbieri:97a} or asymmetric textures~\cite{Babu:95a} are considered. \Eqs{y1} also show that the top and bottom Yukawa couplings are of the same order, i.e.\ $\tan\beta$ is large. This is a prediction of the left-handed dominance scenario, which holds in the absence of significant Higgs mixing. Note also that the simplest way to account for the more pronounced hierarchy in the up quark sector, $m_c/m_t\ll m_s/m_b$ is to have $\epsilon_U\ll\epsilon_D$ and therefore a double hierarchy $M_Q\ll M_D\ll M_U$. We will see below that $m_c/m_t\ll m_s/m_b$ can actually be explained without introducing a third scale. 

The textures in \Eqs{y1} also have an unpleasant feature. Although the masses of the first family fermions have still to be generated, the Cabibbo angle does not vanish and ends up being typically large:
\begin{equation}
\label{eq:Cabibbo}
\tan\theta_C = \fracwithdelims{|}{|}{r^D_1}{r^D_2}\sim 1.
\end{equation}
While the actual value of the Cabibbo angle is not very small and could be accomodated by e.g.\ an accidental cancellation, we prefer to consider its smallness as the indication of a non-accidental correlation between the $\lambda^{qU}_{1,2}$ and $\lambda^{qD}_{1,2}$ coefficients in the initial basis. In turn, such a correlation points at an $\text{SU(2)}_R$ gauge symmetry~\cite{Mohapatra:75a} forcing \globallabel{eq:yukawaLR}
\begin{align}
\lambda^{Qu}_i &= \lambda^{Qd}_i  & \lambda^{Ln}_i &= \lambda^{Le}_i & \alpha^U_i &= \alpha^D_i  \mytag \\
\lambda^{Uq}_i &= \lambda^{Dq}_i & \lambda^{Nl}_i &= \lambda^{El}_i & \alpha^N_i &= \alpha^E_i .\mytag 
\end{align}
We are therefore lead to a $G_{LR} = \text{SU(2)}_L \times \text{SU(2)}_R  \times \text{SU(3)}_c \times \text{U(1)}_{B-L}$ extension of the SM gauge group\footnote{Note that in the presence of an SU(2)$_R$ symmetry the possibility of right-handed dominance also opens up. In fact, the argument leading to left-handed dominance holds under the assumption that the couplings in different sectors, in particular in the right-handed up and down sectors, are uncorrelated. On the other hand, we just saw that the SU(2)$_R$ symmetry does correlate quantities involving right handed up and down quarks and leptons. As a consequence, the possibility that the $Q+\bar Q$ exchange be subdominant to the $Q^c + \bar Q^c$ exchange opens up. In this context, one finds $\lambda_c\sim\lambda_s$ and therefore $\tan\beta\sim m_c/m_s$. The $Q$ and $Q_c$ dominance scenarios are therefore characterized by different predictions for $\tan\beta$. We do not pursue this possibility further in this paper.}. \Eqs{yukawaLR} lead to $\lambda^{Dq}_{1} = 0$, $r^D_1=0$, and therefore $V_{us} = 0$, as anticipated. A non-vanishing value of $V_{us}$ will be generated by the breaking of the SU(2)$_R$ symmetry, which is anyway needed. The standard way to break $G_{LR}$ to $G_\text{SM}$ is through the vev of the scalar component $\tilde L'_c$ ($\tilde {\bar L}'_c$) of a ($\mathbf{Z}_2$-even in our case) chiral right-handed doublet $L'_c$ ($\bar L'_c$) transforming as $L^c = (N^c,E^c)^T$ ($\bar L^c = (\bar N^c,\bar E^c)^T$). 

With the basis choice above, all the first family $\mathbf{Z}_2$-odd fermions have the same charge under the accidental chiral U(1) symmetry protecting the first family, whereas all the other fields are invariant. While a non-vanishing $V_{us}$ will need the breaking of the SU(2)$_R$ symmetry, a non vanishing mass for the first family will need the breaking of that accidental chiral U(1). The accidental family symmetry protecting the second family emerges in the limit in which $U^c$, $D^c$ become heavy so that they can be integrated out. All the second family fermions have the same charge under it. 

\subsection{Neutrino masses and mixing}

We have seen above that small mixing angles are easily obtained in the quark mass sector. At the same time, large mixing angles naturally appear in the neutrino sector provided that the right-handed neutrino messengers $N^c$, $\bar N^c$ dominate the see-saw. This is closely related to the peculiar features of our setting, as we now see. 

As in the quark sector, it is convenient to consider a basis in which  $\alpha^L_{1,2}=\lambda^{Ln}_{1,2} = \lambda^{Le}_{1,2} = 0$ and $\alpha^N_1 = \alpha^E_1 = \lambda^{Nl}_1 = \lambda^{El}_1 = 0$. Because of the left-handed dominance hypothesis, this choice makes in fact the charged lepton Yukawa matrix approximately diagonal. On the other hand, the couplings $\lambda^{Nl}_{2,3}$ of $N^c$ to $l_2$ and $l_3$ are expected to be comparable. We have in fact already used our freedom to redefine $l_2,l_3$ to make the mixings small in the charged lepton sector. As the charged leptons are approximately diagonal, this means that the singlet neutrino $N^c$ has similar $\ord{1}$ couplings to $\nu_\mu$ and $\nu_\tau$. If $N^c$ dominates the see-saw, this is precisely the condition needed to obtain a large atmospheric mixing angle and normal hierarchical neutrino masses in a natural way~\cite{King:98a}. We will see in the next section that all the heavy singlet neutrino masses will be approximately at the same scale, but the ``$N^cN^c$'' entry of the inverse heavy Majorana mass can still dominate the see-saw mechanism. Note that this is an example of see-saw dominated by a singlet neutrino that is not a Pati-Salam (or SO(10)) partner of the light lepton doublets. 

\subsection{The charm quark Yukawa and Pati-Salam}

Since the fields $U^c$ and $D^c$ are unified in a right-handed doublet $Q^c = (U^c, D^c)^T$, an unwanted consequence of the SU(2)$_R$ symmetry is $M_U = M_D = M_{Q^c}$, which gives $m_c/m_t \approx m_s/m_b$. The SU(2)$_R$ symmetry must therefore on the one hand protect $V_{us}$, on the other be badly broken in order to differentiate the charm and strange Yukawas. This apparent problem turns out to provide additional insight on the structure of the model. 

It turns out that an indirect coupling of the available source of SU(2)$_R$ breaking (the scalar fields $\tilde L'_c$, $\tilde {\bar L}'_c$) to the fermions $Q^c$, $\bar Q^c$ is the simplest and most natural way to achieve the hierarchy $m_c/m_t \ll m_s/m_b$. Coupling $(\tilde L'_c,\tilde {\bar L}'_c)$ to $(Q^c,\bar Q^c)$ at the renormalizable level needs the introduction of new fields. There are only two possibilities. The one we are interested in is a vectorlike pair of fermion fields $T+\bar T$ transforming as $(1,1,3,4/3) + (1,1,\bar 3,-4/3)$ under $G_{LR}$ (the last entry denotes the value of $B-L$). Such fields couple to the  $(\tilde L'_c, \tilde {\bar L}'_c)$ and $(Q^c,\bar Q^c)$ fields through the interaction $T Q^c \tilde {\bar L}'_c$ and $\bar T \bar Q^c \tilde {L}'_c$. Once the scalar doublets get a vev, the latter interactions contributes to the masses in the up sector and allows to suppress the charm mass, as we will see in Section~\ref{sec:ren}. The second possibility\footnote{A vectorlike pair $S+\bar S$ transforming as $(1,1,3,-2/3) + (1,1,\bar 3,2/3)$ and coupling through $S Q^c \tilde {L}'_c$ and $\bar S \bar Q^c \tilde {\bar L}'_c$.} does not suppress the charm mass, as it only affects the down quark sector. It can play a role in the case of right-handed dominance. 

The introduction of fermions with the quantum numbers of $T+\bar T$ might look at first sight quite ``ad hoc''. On the other hand, such fermions automatically arise with the Pati-Salam extension of the $G_{LR}$ group, $G_{PS} = \text{SU(4)}_c \times \text{SU(2)}_L \times \text{SU(2)}_R$. The quantum numbers of $T+\bar T$ appear in fact in the decomposition under $G_{LR}$ of the SU(4)$_c$ adjoint and their interactions follow from the standard coupling of the adjoint to the fundamental of SU(4)$_c$. In particular, fields with the quantum numbers of $T+\bar T$ can certainly be found among the SU(4)$_c$ gauginos\footnote{Note that such $T+\bar T$ gauginos automatically get a heavy mass and are thus splitted from the lighter gluinos by the $\text{SU(4)}_c\to\text{SU(3)}_c$ spontaneous breaking. Note also that the required coupling with $Q^c \tilde {\bar L}'$ is also automatically present in the form of a supersymmetric gauge interaction, provided that $\tilde L'$ is the partner of $L$.}. Unfortunately the simplest implementation of the economical interpretation in which the $T+\bar T$ fields are gauginos and $L'=L$ leads to problems in the Higgs sector. In order to avoid those problem we will make sure that $R$-parity is not broken, which requires $T+\bar T$ and $\tilde L'$, $\tilde {\bar L}'_c$ to be associated to new chiral fields. 

\section{A model of flavour from accidental symmetries}
\label{sec:analysis}

\subsection{Definition of the model}
\label{sec:definition}

The chiral superfield content of the model and the quantum numbers under $G_{\text{PS}}$ and $\mathbf{Z}_2$ are specified in Table~\ref{tab:fields}. The first block contains the $\mathbf{Z}_2$-odd fields: the 3 light (in the unbroken $\mathbf{Z}_2$ limit) families $(f_i,f^c_i)$, $i=1,2,3$, the light Higgs $h$ and the $\mathbf{Z}_2$-breaking field $\phi$. The latter is in the adjoint representation of SU(4)$_c$ as this provides the Georgi-Jarlskog factor 3 needed to account for the $\mu$--$s$ mass relation. The second block contains the messengers, in a single vectorlike family $(F,F_c)+(\bar F,\bar F_c)$. A Higgs messenger is also included, corresponding to \Fig{diagrams}c. The third block contains the fields $F'_c+\bar F'_c$ breaking the Pati-Salam group (including the SU(2)$_R$ subgroup)  and an $\mathbf{Z}_2$-even SU(4)$_c$ adjoint $\Sigma$ providing the fields $T+\bar T$ discussed in Section~\ref{sec:motivations}. SO(10) partners $F'+\bar F'$ of $F'_c+\bar F'_c$ are also included. The last block contains two sources of Pati-Salam breaking. They contain the two possible SM invariant directions in the Pati-Salam adjoint. Table~\ref{tab:fields} also shows the $R$-parity associated to each field. $R$-parity plays a role in preventing the economical identification of the primed fields with $F^c$ and $\bar F^c$ and of $\Sigma$ with the SU(4)$_c$ gauginos. When discussing the neutrino sector we will also introduce Pati-Salam singlets. 

\begin{table}[htdp]
\begin{equation*}
\begin{array}{|c|cccc|ccccc|ccccc|cc|}
\hline
& f_i & f^c_i & h & \phi & F & \bar F & F^c & \bar F^c & H & F' & \bar F' & F'_c &  \bar F'_c & \Sigma & X & X_c \\ \hline
\text{SU(2)}_L & 2 & 1 & 2 & 1 & 2 & 2 & 1 & 1 & 2  & 2 & 2 & 1 & 1& 1 & 1 & 1 \\
\text{SU(2)}_R & 1 & 2 & 2 & 1 & 1 & 1 & 2 & 2 & 2  & 1 & 1 & 2 & 2 & 1 & 1 & 3 \\
\text{SU(4)}_c & 4 & \bar 4 &  1 & 15 & 4 & \bar 4 & \bar 4 & 4 & 1 & 4 & \bar 4 & \bar 4 & 4 & 15 & 15 & 1 \\
\mathbf{Z}_2 & - & - & - & - & + & + & + & + & + & + & + & + & + & + & + & + \\
R_P & - & - & + & + & - & - & - & - & + & + & + & + & + & - & + & + \\
\hline
\end{array}
\end{equation*}
\caption{Field content of the model and quantum numbers under $G_{\text{PS}}$ and $\mathbf{Z}_2$}
\label{tab:fields}
\end{table}

Our hypothesis is that the Pati-Salam gauge structure and the fields in Table~\ref{tab:fields} happen to be the only relatively light fields surviving below the cutoff $\Lambda$ of our theory, which will not be very far from $10^{16}\GeV$. We implement this hypothesis by linking the mass of the heavy fields to Pati-Salam breaking. We do not address the origin of this assumption here. No dynamics related to the family indices is required. On the contrary, we will assume that the dimensionless coefficients in the superpotential are $\ord{1}$ and uncorrelated. 

The renormalizable part of the superpotential is
\begin{multline}
\label{eq:RenLag}
W^\text{ren} =
\lambda_i f^c_i F h + \lambda^c_i f_i F^c h +\alpha_i\phi f_i\bar F+\alpha^c_i\phi f^c_i\bar F^c + X\bar F F + X_c \bar F^c F^c   \\
 + \bar\sigma_c \bar F'_c \Sigma F^c +\sigma_c\bar F^c\Sigma F'_c + \bar\sigma \bar F' \Sigma F +\sigma \bar F \Sigma F' +\gamma X\Sigma^2 
 % + W^{\text{ren}}_\text{vevs}(X, X_c, \phi, F'_c, \bar F'_c, F', \bar F') 
 \\
 + \lambda^H_{ij} f^c_i f_j H + \eta F^c F H + \bar\eta \bar F^c \bar F H + \eta' F'_c F' H + \bar\eta' \bar F'_c \bar F' H . 
\end{multline}
We have included all terms compatible with our hypotheses except a mass term for the Higgses $h$ and $H$. We have not shown the part of the superpotential involving the primed fields and all other fields getting a vev. An irrelevant term $X\bar F^c F^c$ is also omitted. As anticipated, the messenger fields and $\Sigma$ only get a mass through the Pati-Salam breaking fields. Besides $X$, $X_c$, the fields getting a vev are $\phi$, $F'_c$, $\bar F'_c$ ($R_P$ is thus preserved). The hierarchy of fermion masses originates from the assumption that the Pati-Salam breakings along the $T_{3R}$ and $N'_c,\bar N'_c$ directions, $\vev{X_c} = M_c (2T_{3R})$ and $\vev{F'_c} = (V_c, 0)^T$, $\vev{\bar F'_c} = (\bar V_c, 0)^T$ respectively, both take place at a scale $M_c\sim V_c$ much higher scale than the scale $M\sim v$ of the breaking along the $B-L$ direction, $\vev{X} = M T_{B-L}$, $\vev{\phi} = v T_{B-L}$.\footnote{One example for the superpotential involving the primed fields and $X_c,X,\phi$ only is (neglecting $F'$, $\bar F'$, including mass terms)
\[
W' = (M_c -\delta_c X_c)\bar F'_c F'_c + \frac{M_{X_c}}{2}X_c^2 + \frac{M_X}{2}X^2 +\frac{M_\phi}{2}\phi^2 + \rho_1 X^3 +\rho_2 X\phi^2 .
\]
This it the most general renormalizable potential except for the $X \bar F'_c F'_c$ coupling, which is assumed to vanish. One solution of the $F$-term equations is (up to an SU(2)$_R$ rotation) $\delta_c \vev{X} = M_c (2T_{3R})$, $(\delta_c/2)^2 \vev{\bar N'_c N'_c} = M_{X_c}^2$, $\vev{\phi}=0$, $\vev{X} = 0$. Both the breaking along the $T_{3R}$ and $N'_c,\bar N'_c$ directions take place at the same scale $M_c$, while the breaking along the $B-L$ direction is suppressed (zero at the renormalizable level). 
% Note that the ``sliding singlet'' mechanism at work makes a Dirac singlet neutrino massless.
} The horizontal fermion hierarchy therefore follows from the vertical structure of the theory. The vev of $\phi$ breaks the $\mathbf{Z}_2$ symmetry and mixes light and heavy fields, thus giving rise to the Yukawa couplings of light fields. The vevs of $F'_c$ and $\bar F'_c$ are responsible for the full breaking of the Pati-Salam to the SM group, they generate a mixing between SU(3)$_c$ triplets which suppresses the charm quark Yukawa, and they make $H$ heavy. 

It is convenient to choose a basis in flavour space such that $\lambda_{1,2} = \alpha_{1,2}=0$, $\lambda^c_1 = \alpha^c_1 = 0$. Moreover, $\lambda_3$, $\alpha_3$, $\lambda^c_{2,3}$, $\alpha^c_{2,3}$, $\gamma$, $M$, $M^c$, $\bar\sigma_c$, $\vev{\phi}$, $V_c = \bar V_c$,  can all be taken positive. We therefore see that the effective theory in which $H$ is integrated out possesses an accidental chiral U(1)$_1$ flavour symmetry protecting the first family Yukawas: $f_1 \to e^{i\alpha} f_1$, $f^c_1 \to e^{i\alpha} f^c_1$. In the limit in which the heavier messengers $F^c,\bar F^c$ are also integrated out, an additional accidental flavour symmetry U(1)$_2$ protects the second family Yukawas: $f_2 \to e^{i\beta} f_2$, $f^c_2 \to e^{i\beta} f^c_2$. The hierarchy between the third and the first two fermion family masses can be seen as a consequence of the above flavour symmetries. The stronger suppression of the first fermion family mass is due to the fact that the heavy Higgs $H$ does not mix with $h$ at the renormalizable level. This is because the coupling $\phi H h$ is not allowed by the SU(4)$_c$ symmetry. The suppression of the first family masses is therefore obtained for free, as it is a consequence of the Pati-Salam quantum numbers of $\phi$, which are independently motivated by the $m_\mu/m_s$ ratio. 

\subsection{The fermion spectrum at the renormalizable level}
\label{sec:ren}

Since $R$-parity is not broken, we can confine ourselves to the $R_P$-odd fields. Let us denote by $A_\Sigma$, $T_\Sigma$, $\bar T_\Sigma$, $G_\Sigma$ the (properly normalized) SM components of $\Sigma$. Under $\text{SU(3)}_c\times \text{SU(2)}_w \times\text{U(1)}_Y$, $A$ is a singlet, $T\sim(3,1,2/3)$ is a color triplet, $\bar T\sim (\bar 3, 1, -2/3)$ is an antitriplet, $G\sim (8,1,1)$ is an octet. With standard notations for the SM components of the fields in Table~\ref{tab:fields}, the mass terms are
\begin{equation}
\label{eq:masses}
\begin{gathered}
-\bar L \left[ M L +\alpha_3 v l_3 \right] - 
\bar E^c \left[ M_c E^c + v(\alpha^c_3 e^c_3 +\alpha^c_2 e^c_2) \right]  \\
+ \frac{1}{3}\bar Q \left[ M Q +\alpha_3 v q_3 \right] - 
\bar D^c \left[ M_c D^c -\frac{v}{3}(\alpha^c_3 d^c_3 +\alpha^c_2 d^c_2) \right] \\
+\bar U^c \left[ M_c U^c +\frac{\sigma_c}{\sqrt{2}}  V_c\bar T_\Sigma +\frac{v}{3} (\alpha^c_3 u^c_3 + \alpha^c_2 u^c_2) \right]
+T_\Sigma \left[M_\Sigma \bar T_\Sigma +\frac{\bar\sigma_c}{\sqrt{2}} %\bar 
V_c U^c
\right]  \\
+ \bar N^c \left[M_c N^c -v(\alpha^c_3 n^c_3 +\alpha^c_2 n^c_2) \right]  
- \sqrt{\frac{3}{8}} \sigma_c V^c \bar N^c A_\Sigma - \sqrt{\frac{3}{8}} \bar\sigma_c %\bar 
V_c N^c A_\Sigma +M_\Sigma A^2_\Sigma  \\
+ \eta'V_c L' H_u + \bar\eta' V_c \bar L' H_d
- \frac{M_\Sigma}{2} G^2_\Sigma , 
\end{gathered}
\end{equation}
where $M_\Sigma = -(2/3) \gamma M$. The charged fermion Yukawas are obtained by identifying the massless combinations and expressing the Yukawa lagrangian 
\begin{equation}
\lambda^c_i U^cq_ih_u + \lambda^c_i D^cq_ih_d + \lambda^c_i N^cl_ih_u + \lambda^c_i E^cl_ih_d +
\lambda_i u^c_i Q h_u +\lambda_i d^c_i Q h_d +\lambda_i n^c_i L h_u +\lambda_i e^c_i L h_d 
\end{equation}
in terms of them. We then obtain, at the scale $M$ and at the leading order in $\epsilon$, 
\begin{equation}
\label{eq:YDE}
Y^D =  \begin{pmatrix}
0 & 0 & 0 \\
0 & \alpha^c_2\lambda^c_2 \epsilon/3 & \alpha^c_2\lambda^c_3 c\, \epsilon/3 \\
0 & \alpha^c_3\lambda^c_2 \epsilon/3 & -s\lambda_3
\end{pmatrix} \qquad
Y^E = -\begin{pmatrix}
0 & 0 & 0 \\
0 & \alpha^c_2\lambda^c_2 \epsilon & \alpha^c_2\lambda^c_3 c\, \epsilon \\
0 & \alpha^c_3\lambda^c_2 \epsilon & s\lambda_3
\end{pmatrix} ,
\end{equation}
where $c=\cos\theta$, $s=\sin\theta$, $\tan\theta \equiv \alpha_3 v/M =\ord{1}$, $\epsilon \equiv v/M_c \ll 1$. The numerical value of $\epsilon$ turns out to be $\epsilon \approx 0.06\,(s\lambda_3)/(\alpha^c_2\lambda^c_2)$. 

The up quark sector deserves some additional comments. The situation is different than in the down quark and charged lepton sector, as the triplet $\bar T_\Sigma$ has the same SM quantum numbers as $u^c_i$ and $U^c$ and mixes as well. The charm quark Yukawa arises from the interaction $\lambda^c_i U^c q_i h_U$ when $U^c$ is replaced by its light component. The light component must be orthogonal to both the combinations in squared brackets in the third line of \eq{masses}. As a consequence, the charm Yukawa turns out to be suppressed twice. The light component of $U^c$ vanishes in fact both in the $v\to0$ limit ($\mathbf{Z}_2$ is not broken, $u^c_i$ do not mix with $U^c,\bar T_\Sigma$) and in the $M_\Sigma\to 0$ limit (the light component must in this case be orthogonal to $U^c$). This explains the factors $\epsilon^2$ in 
\begin{equation}
\label{eq:YU}
Y^U = -\begin{pmatrix}
0 & 0 & 0 \\
0 & (4/9)\alpha^c_2\lambda^c_2 \rho_u\epsilon^2 & (4/9)\alpha^c_2\lambda^c_3 c \rho_u\epsilon^2 \\
0 & (4/9)\alpha^c_3\lambda^c_2 \rho_u\epsilon^2 & s\lambda_3
\end{pmatrix} .
\end{equation}
In the equation above, $\rho_u = (\gamma\alpha_3)/(\sigma_c\bar\sigma_c t_\theta)(M_c/V_c)^2$, which turns out to be close to one as it should, as $\rho_u\,\epsilon\approx 0.07$--$0.08$. 

The Yukawas of the first family vanish at the renormalizable level, as anticipated. We will see below how they are generated at the non-renormalizable level. For the time being, let us comment about some interesting features of eqs.~(\ref{eq:YDE},\ref{eq:YU}). We have assumed that i) the $\mathbf{Z}_2$-breaking field $\phi$ is in the adjoint of SU(4)$_c$ and ii) the masses of the messenger fields and $\Sigma$ are linked to Pati-Salam breaking, with the breaking along the $B-L$ direction taking place at a much smaller scale than the breaking in the $T_{3R}$ and singlet neutrino directions. As a consequence, we find i) $m_s\ll m_b$ and $m_\mu\ll m_\tau$, ii) $|V_{cb}|\sim m_s/m_b$, iii) $(m_\tau/m_b)_M \approx 1$ iv) $(m_\mu/m_s)_M \approx 3$, v) $m_c/m_t\ll m_s/m_b$. We also predict the suppression of the first family fermion masses. Note in particular that two different hierarchies in the down quark/charged lepton sectors and in the up quark sector are obtained in terms of a single hierarchy between the two scales of the theory $M_c$ and $M$. Note also that the relation $|V_{cb}|\sim m_s/m_b$ is a direct consequence of the principles of our approach. As usual in the presence of a single Higgs multiplet,  one also obtains $\lambda_\tau$--$\lambda_b$--$\lambda_t$ unification. 

Let us now consider the neutrino sector. The ($R_P$-odd) SM singlet neutrino fields in the model are $n^c_{1,2,3}$, $N^c$, $\bar N^c$, $A_\Sigma$. \Eq{masses} shows that $\alpha^c_3 n^c_3 +\alpha^c_2 n^c_2$, $N^c$, $\bar N^c$, $A_\Sigma$ get a heavy mass, while $\alpha^c_2 n^c_3-\alpha^c_3 n^c_2$ and $n^c_1$ are massless at the renormalizable level. This is clearly a problem, as it implies a Dirac mass to the tau neutrino at the electroweak scale. A possible solution is to invoke (small) non-renormalizable contributions to the massesÄ of $\alpha^c_2 n^c_3-\alpha^c_3 n^c_2$ and $n^c_1$. However, this would make the latter fields dominate the see-saw, while we saw in the previous section that we prefer $N^c$ to dominate. We therefore couple the SM singlets $n^c_i$ to 3 Pati-Salam singlets $s_i\sim (1,1,1,-,-)$ through the Dirac mass term provided by the interaction $\eta^s_{ki}s_k f^c_i \bar F'_c$. This raises the fields $n^c_i$ and $s_k$ at the higher of the two scales of our model. Note that it is always possible to choose a basis for the $s_k$'s such that the coupling $\eta^s_{ki}$ and the Dirac mass term are diagonal.

The fields $n^c_i$ and $s_k$ constitute a pseudo-Dirac system. That is because a Pati-Salam invariant Majorana mass term for the Pati-Salam singlets $s_k$ cannot be written at the renormalizable level, according to our hypothesis stating that the mass terms should originate from PS breaking. The only correction to the pure Dirac limit therefore comes from the mixing of the $s_k$'s with $A_\Sigma$, which is however suppressed by $v/M_c = \epsilon$. Since the coupling of the pseudo-Dirac pair $(n^c_3,s_3)$, to the light lepton doublets, $\lambda_3 n^c_3 L h_u$, only involves $n^c_3$, the contribution to the see-saw of the $(n^c_i,s_i)$ fields is negligible. In fact, that contribution vanishes in the pure Dirac limit. This can be seen for example by diagonalizing the Dirac pairs in terms of two Majorana mass eigenstates with opposite mass. As in the Dirac limit $n_3$ contains the two eigenstates with exactly the same weight, the two contributions to the see-saw exactly cancel\footnote{An alternative way to verify that the Dirac system does not contribute to the see-saw is to observe that its contribution is proportional to $(M_D^{-1})_{n^c_3n^c_3}$, where $M_D$ is the Dirac mass term for the two Weyl spinors $n^c_3$, $s_3$ with vanishing diagonal entries. As the inverse of a Dirac mass matrix is still in the Dirac form, $(M_D^{-1})_{n^c_3n^c_3} = 0$}. Taking into account the small corrections to the pure Dirac limit, the contribution of $(n^c_i,s_i)$ to the see-saw turns out to be suppressed by $\epsilon$. More precisely, the contribution to the atmospheric angle is suppressed by $\epsilon$ and the contribution to $m_2/m_3$ by $\epsilon^2$. We can then safely neglect the fields $n^c_i$ and $s_k$ for our purposes. This can also be verified by using the full $9\times 9$ singlet neutrino mass matrix in the see-saw formula. 

We are then left with 3 SM singlet (right-handed) neutrinos $N^c$, $\bar N^c$, $A_\Sigma$ with mass terms
\begin{equation}
\label{eq:numasses}
M^c\bar N^c N^c -\sqrt{\frac{3}{8}} V_c A_\Sigma (\sigma_c\bar N^c +\bar\sigma_c N^c) + M_\Sigma A^2_\Sigma
\end{equation}
entering the see-saw through the Yukawa interaction $N^c (\lambda^c_3l_3+\lambda^c_2l_2)h_u$. The following effective $D=5$ left-handed neutrino mass operator is then generated
\begin{equation}
\label{eq:seesaw}
\frac{1}{4}\frac{\sigma_c}{\bar\sigma_c} \frac{1}{M_c} (c\lambda^c_3l'_3+\lambda^c_2 l'_2)^2  h^2_u ,
\end{equation}
where $l'_3 = c l_3 -s L$, $l'_2 = l_2$ are the light lepton doublets. We have therefore obtained a normal hierarchy and a large atmospheric mixing angle $\theta_{23}$ in a natural way,
\begin{equation}
\label{eq:atm}
\tan\theta_{23} = \frac{\lambda^c_2}{c\lambda^c_3}, \qquad m_3 = \rho_\nu \frac{v_\text{EW}^2}{2s^2_{23}M_c}, \qquad m_{1,2} \approx 0,
\end{equation}
where $v_{\text{EW}}\approx 174\GeV$ is the electroweak breaking scale, $s_{23} = \sin\theta_{23}$, and $\rho_\nu = (\sigma_c/\bar\sigma_c)(\lambda^c_2)^2 \sim 1$. \Eq{atm} determines the scale $M_c$ of our model, $M_c\approx 0.6\cdot 10^{15}\GeV\rho_\nu$. The solar mixing angle and mass difference are generated at the non-renormalizable level together with the masses of the first charged fermion masses. 

\subsection{The first family}

As discussed, the first family fermion masses are protected by an accidental U(1)$_1$ family symmetry. That symmetry is actually broken by the coupling of the first family with the heavy Higgs messenger $H$. However, $H$ does not mix with the light Higgs $h$ at the renormalizable level, which means that for our purposes it is effectively decoupled. The U(1)$_1$ symmetry can therefore be broken by non-renormalizable interactions either because the interactions directly involve the first family or because they induce a $H$-$h$ mixing. Here we will consider the second possibility. In both cases, the first family mass will be further suppressed with respect to the other families by the heavy cutoff scale $\Lambda$. 

Not all the non-renormalizable operators are suitable to give a mass to the first family. For example, the operator $f^c_i f_j \phi h$ gives the same contribution to the Yukawas of the up and down quarks (in this $\lambda_t \approx \lambda_b$ scenario the up quark mass Yukawa needs to be suppressed by a factor of about 200). The operator $F'_c F' \phi h$ is also dangerous, as it indirectly contributes to the up quark mass only. We therefore need to make an assumption on the operators generated by the physics above the cutoff $\Lambda$. A simple assumption is that the the heavy physics only couples $\phi$ to the barred $\bar F'$, $\bar F'_c$ (but not to $F'$, $F'_c$). This would still allow an operator in the form 
\begin{equation}
\label{eq:NRfirstfamily}
\frac{a}{\Lambda} \bar F'_c \bar F' \phi h ,
\end{equation}
which turns out to give mass to the electron and the down quark, but not to the up quark, as desired. The reason is that the operator above induces a mixing in the down Higgs sector but not in the up Higgs sector. As mentioned in Section~\ref{sec:definition}, $H_d$ and $H_u$ get a mass term, $\eta' V_c L' H_u + \bar\eta' V_c \bar L' H_d$, from the vev of $\bar F'_c$ through the renormalizable interactions in \eq{RenLag}. In addition, the operator in \eq{NRfirstfamily} gives a mass term $-a(V_c v/\Lambda) \bar L' h_d$, which induces a mixing between the two down Higgses $H_d$ and $h_d$. This in turn communicates the U(1)$_1$ breaking provided by $\lambda^H_{ij} f^c_i f_j H$ to the down quark and charged lepton sector. When $H_d$ is expressed in terms of the exact Higgs mass eigestates $H'_d$ and $h'_d$, the latter operator induces in fact a contribution  to the down and charged lepton Yukawas matrices $Y^D_{ij}$ and $Y^E_{ij}$ given by $\epsilon'\rho_h\lambda^H_{ij}$ (up to the $L$-$l'_3$ mixing), where
\begin{equation}
\label{eq:eps1}
\epsilon' = \frac{v}{\Lambda} = \epsilon \frac{M_c}{\Lambda} 
\end{equation}
and $\rho_h = a/\bar\eta'\sim 1$. The small ratio $M_c/\Lambda$ explains the further suppression of the first fermion family. We then obtain, at leading order,
\begin{equation}
\label{eq:YDE2}
Y^D =  \begin{pmatrix}
\rho_h \lambda^H_{11} \epsilon' & \rho_h \lambda^H_{12} \epsilon' & \rho_h \lambda^H_{13} c\, \epsilon' \\
\rho_h \lambda^H_{21} \epsilon' & \alpha^c_2\lambda^c_2 \epsilon/3 & \alpha^c_2\lambda^c_3 c\, \epsilon/3 \\
\rho_h \lambda^H_{31} \epsilon' & \alpha^c_3\lambda^c_2 \epsilon/3 & -s\lambda_3
\end{pmatrix} \qquad
Y^E = \begin{pmatrix}
\rho_h \lambda^H_{11} \epsilon' & \rho_h \lambda^H_{12} \epsilon' & \rho_h \lambda^H_{13} c\, \epsilon' \\
\rho_h \lambda^H_{21} \epsilon' & -\alpha^c_2\lambda^c_2 \epsilon & -\alpha^c_2\lambda^c_3 c\, \epsilon \\
\rho_h \lambda^H_{31} \epsilon' & -\alpha^c_3\lambda^c_2 \epsilon & -s\lambda_3
\end{pmatrix} .
\end{equation}
The up Higgs does not mix, which explains the smallness of the up quark Yukawa. The latter will be eventually generated by Planck scale effects. For example an operator $(c/M_{\text{pl}}) f^c_i f_j \phi h$ would provide a up quark Yukawa of the correct order of magnitude for $c\sim 1$. The latter argument also provides an independent estimate (an upper bound in the general case) of the scale $M_c$, which happens to coincide with our estimate from neutrino physics. 

\Eq{YDE2} shows that the electron and down quark masses are expected to be similar, while the correct relation is $m_e\sim m_d/3$ at the heavy scale. In order to avoid the wrong relation, $\lambda^H_{11}$ should be sufficiently suppressed in the basis in flavour space which identifies the first family. Quantitatively, the requirement is $\lambda^H_{11}/\lambda^H_{12,21} < \sqrt{m_d/m_s}/3\sim 0.08$. This suppression could for example accidentally arise when rotating the fields to go in the basis in which eqs.~(\ref{eq:YDE},\ref{eq:YDE2}) are written. In this case one obtains $m_e\sim m_d/3$ and $V_{us}\sim \sqrt{m_d/m_s}$, as observed, at the price of a fine-tuning of at least $\mathcal{O}(10)$\footnote{One could make at this point the totally disinterested observation that our model involves more than $\mathcal{O}(10)$ relations among $\ord{1}$ coefficients, so that accidental cancellation of leaving less than one part out of 10 is expected to occur somewhere. In fact, from this point of view, the distribution of the absolute values of our $\ord{1}$ coefficients turns out to be rather peaked on 1.}.

The full CKM matrix can be obtained by diagonalizing the up and down Yukawa matrices. $V_{ub}/V_{cb}$ and $V_{td}/V_{ts}$ both get a contribution from $Y^D_{31}$. On top of that, $V_{td}/V_{ts}$ also gets a contribution from the commutation of the ``12'' rotation used to diagonalize $Y^D$ and the relative 23 rotation($V_{cb}$). In formulas, 
\begin{equation}
\label{eq:CKM}
\frac{V_{ub}}{V_{cb}} = \frac{\alpha^c_2\lambda^H_{31}}{\alpha^c_3\lambda^H_{21}} V_{us}, \qquad
\delta = \arg\left[ \frac{\alpha^c_2\lambda^H_{31}}{\alpha^c_3\lambda^H_{21}}\right], \qquad
\left|\frac{V_{td}}{V_{ts}}\right| = \left| |V_{us}| - \left|\frac{V_{ub}}{V_{cb}}\right| e^{i\delta}\right| ,
\end{equation}
where $\delta$ is the CKM phase in the standard parameterization. The present SM CKM fits give~\cite{UTfit} $|(\alpha^c_2\lambda^H_{31})/(\alpha^c_3\lambda^H_{21})|\approx 0.4$.

A comment on $V_{us}$ is in order. As we saw, the physics giving rise to the Yukawas of the first family will typically also generate a contribution to $V_{us}$. $V_{us}$ and the first family are however in principle independent issues. In fact, $V_{us}$ is related to the breaking of the LR symmetry, while the first family requires the breaking of the corresponding accidental flavour symmetry. Indeed, the reason why the mechanism generating first family Yukawas also typically generates $V_{us}$ is that in order to make $m_d/m_b\gg m_u/m_t$ the LR symmetry must be broken. On the other hand, it is possible to generate a contribution to $V_{us}$ without inducing a corresponding contribution to the first family mass. The operator $b_i X_c F^c f_i h/\Lambda$, involving the SU(2)$_R$ breaking field $X_c$, gives for example a contribution $ 2(b_1/\lambda^c_2) (M_c/\Lambda)$ to $V_{us}$ without breaking U(1)$_1$ (it also modifies \eq{CKM}). From the previous argument and from \eq{YDE2} we expect 
\begin{equation}
\label{eq:Vus}
\frac{M_c}{\Lambda}\sim \frac{|V_{us}|}{2}\sim 0.1 .
\end{equation}

Finally, let us go back to neutrino masses. By using the renormalizable interactions, we succeeded in giving a mass to the heaviest neutrino $\nu_3$ and in generating a large atmospheric neutrino angle $\theta_{23}$. We still need to generate a mass for the intermediate neutrino $m_2$ and a corresponding large solar angle $\theta_{12}$. As shown in the Appendix, non-renormalisable interactions involving the fields introduced so far can generate a mass term for $m_2$ at the correct level together with a non-vanishing $\theta_{13}$ close to the current experimental limit, but not a large solar angle $\theta_{12}$. However, a large solar angle can be induced by a Pati-Salam singlet $S\sim (1,1,1,+,-)$ coupling at the non-renormalizable level only\footnote{This is an important assumption as renormalizable interactions $S\bar F'_cF^c$, $S\bar F^cF'_c$ would in principle be allowed by the symmetries of the theory.}. Its mass term will be in the form $d'(V^2_c/\Lambda)S^2$. Its Yukawa coupling to the lepton doublets comes from the operator $e_iF'_cSf_ih_u/\Lambda$. Its mixing with the other SM singlets is negligible. Therefore, its contribution to the neutrino mass operator is simply given by
\begin{equation}
\label{eq:seesaw4}
-\frac{1}{4d'}\frac{1}{\Lambda} (e_3 c\, l'_3 + e_2 l'_2 +e_1 l'_1)^2  h^2_u .
\end{equation}
We then get an additional contribution to $\theta_{13}$, $\theta_{13}^e = -s^2_{23}\rho_{12} e_1(c\, c_{23} e_3+s_{23}e_2)(M_c/\Lambda)$, where $\rho_{12} = 1/(\rho_\nu d')$. Moreover, in the limit in which only \eq{seesaw4} adds to the leading term in \eq{seesaw}, the lighter neutrino masses $m_1$ and $m_2$, together with the solar mixing angle, are given by the diagonalization of the ``12'' mass matrix
\begin{equation}
\label{eq:m12}
-s^2_{23}\rho_{12} m_3 \frac{M_c}{\Lambda}
\begin{pmatrix}
e^2_1 & e_1(c_{23} e_2-c\,s_{23} e_3) \\
e_1(c_{23} e_2-c\,s_{23} e_3) & (c_{23} e_2-c\,s_{23} e_3)^2
\end{pmatrix} .
\end{equation}

\section{Summary}

In this paper we have proposed a new approach to fermion masses an mixings in which the dominance of a single family of messengers accounts for the lightness of the first family, and the further dominance of the left-handed doublet messengers accounts for the lightness of the second family. With only these assumptions we are able to account for the fermion mass hierarchy, as well as the successful mass relation $m_s/m_b\approx |V_{cb}|$. In order to naturally acount for a small Cabibbo angle, and the correct charm quark mass, we were then led to consider a broken Pati-Salam gauge structure.

The hypothesis underlying our setting is that the Pati-Salam gauge structure, the three SM families, and a relatively small set of heavy fields happen to be the only structure surviving below the cutoff $\Lambda \sim 10^{16- 17}\GeV$ of our model. The flavour structure of the SM fermions essentially only follows from this hypothesis, with no dynamics related to the family indices or detailed knowledge of the theory above the cutoff required. 

This framework has several interesting features. The horizontal hierarchy among different families follows from a vertical hierarchy among messengers belonging to the same family. The latter is in turn related to the breaking pattern of the Pati-Salam group, with the breaking along the $T_{3R}$ and singlet neutrino directions taking place at a higher scale than the breaking along the $B-L$ direction. 
In spite of the absence of small coefficients, the CKM mixing angles turn out to be small. At the same time, a large atmospheric mixing appears in the neutrino sector between normal hierarchical neutrinos in a natural way. This is obtained through a see-saw mechanism dominated by a singlet neutrino $N^c$ which is not unified with the light lepton doublets, as it belongs to the messenger families. The final scheme has $N^c$ as the dominant singlet, with $S$ as the leading subdominant singlet as in sequential dominance. The relation $|V_{cb}|\sim m_s/m_b$ is a direct consequence of the principles of our approach. The two different mass hierarchies in the down quark/charged lepton sectors on one side and in the up quark sector on the other are obtained in terms of a single hierarchy between the two scales of the theory $M_c$ and $M$. The suppression of the first fermion family masses also does not need a new scale for the messenger fields. It is actually a prediction of the model, as it again follows from the gauge structure of the model, which forbids the relevant coupling of the Higgs messenger field. As usual in the presence of a single Higgs multiplet,  one also obtains $\lambda_\tau$--$\lambda_b$--$\lambda_t$ unification. 

The precise structure of the masses and mixings of the first fermion family requires an assumption on the operators generated by the physics above the cutoff $\Lambda$ and relies on an accidental cancellation corresponding to a fine-tuning of at least 10. In the neutrino sector, a large solar mixing angle is obtained together with $\theta_{13} =\ord{m_2/m_3}$, close to the present experimental limit. 

In conclusion, we have proposed the notion of flavour from accidental symmetries as a novel and promising approach to understanding the origin of flavour.

\section*{Acknowledgments}

We thank Graham Ross for useful discussions during the early stages of this work and the CERN Theoretical Physics Unit for its hospitality. A.R. also thanks the Galileo Galilei Institute for Theoretical Physics for the hospitality and the INFN for partial support. 

\section*{Appendix}

In this Appendix we show that in the absence of $S$ non-renormalizable contributions to the superpotential generate a non-vanishing $m_2$ and a sizable contribution to $\theta_{13}$, but no large solar mixing angle. In general, the latter contributions can affect the see-saw either through the singlet neutrino mass matrix or through the Yukawa interactions with the light SM lepton doublets. The leading order operators contributing to the singlet neutrino mass matrix are $\bar F'_c \bar F'_c f^c_i f^c_j$, $\bar F'_c \bar F'_c F^c F^c$, $F'_c F'_c \bar F^c \bar F^c$, $\bar F'_c F'_c s_k s_h$, $X^2_c s_k s_h$. Only the two operators involving $s_k$ affect the see-saw in a significant way. Let $d_{ij} (V^2_c/\Lambda) s_is_j$ be the Majorana mass term induced by those operators. If $M_s$ is the singlet neutrino mass matrix, the $s_3s_3$ mass term gives $(M^{-1}_s)_{n^c_3n^c_3}\approx -2(d_{33}/{\eta^s_3}^2)/\Lambda$. In turn, through the Yukawa interaction $\lambda_3 n^c_3 L h_u$ and the see-saw mechanism, the latter gives a contribution 
\begin{equation}
\label{eq:nu3}
\frac{d}{{\eta^s_3}^2}\frac{1}{\Lambda} (s\lambda_3 l'_3)^2h_u^2
\end{equation}
to the dimension 5 neutrino mass operator, which adds to the leading order contribution in \eq{seesaw}. By diagonalizing the resulting light neutrino mass matrix we then get 
\begin{equation}
\label{eq:m2}
\frac{m_2}{m_3} \approx 
4\rho_{23}\sin^4\theta_{23} \frac{M_c}{\Lambda} ,
\end{equation}
where $\rho_{23} = (s\lambda_3/\lambda^c_2)^2(\bar\sigma_c d)/(\sigma^c{\eta^s_3}^2)\sim 1$ and $\theta_{23}$ is the atmospheric mixing angle. The ratio $m_2/m_3$ turns out to be of the correct order of magnitude given the estimate in \eq{Vus}. 

We also have non-renormalizable contributions to the Yukawa interactions with the light SM lepton doublets. The relevant operators are $b_i X_c F^c f_i h/\Lambda$ and $b'_i\Sigma F'_c f_i h_u/\Lambda$, other possibilities leading to a higher $\epsilon$ suppression. Both operators lead to a contribution to $\theta_{13}$ without inducing a significant solar mixing angle or $m_2/m_3$. We have already discussed the first operator in connection to SU(2)$_R$ breaking and $V_{us}$. In the lepton sector its role is again to misalign the Yukawa couplings of $N^c$ and $E^c$ to the lepton doublets $l_i$. In a basis in which $E^c$ has no Yukawa interaction with $l_1$, the Yukawa interaction of $N^c$ becomes $N^c[\lambda^c_3 l_3 +\lambda^c_2 l_2 +2b_1(M_c/\Lambda) l_1]h_u$ and \eq{seesaw} becomes
\begin{equation}
\label{eq:seesaw2}
\frac{1}{4}\frac{\sigma_c}{\bar\sigma_c} \frac{1}{M_c} \left(c\lambda^c_3l'_3+\lambda^c_2 l'_2 +2b_1\frac{M_c}{\Lambda} l'_1\right)^2  h^2_u .
\end{equation}
The second operator $b'_i\Sigma F'_c f_i h_u/\Lambda$ gives rise to a Yukawa interaction for the singlet $A_\Sigma$, $-\sqrt{3/8}b'_i(V_c/\Lambda) A_\Sigma l_ih_u$, which induces new contributions to the see-saw. In terms of the inverse mass matrix $M^{-1}_s$ of the singlet neutrinos $N^c$, $\bar N^c$, $A_\Sigma$, and in the limit in which the $n^c_i$ contribution is neglected, the neutrino mass operator is in fact now given by 
\begin{multline*}
\frac{1}{2}\left[(M^{-1}_s)_{N^cN^c} (\lambda^c_il_i)^2
+ (M^{-1}_s)_{A^\Sigma A^\Sigma} \left(\sqrt{\frac{3}{8}}b'_i \frac{V^c}{\Lambda}l_i\right)^2
- 2(M^{-1}_s)_{A^\Sigma N^c}  \left(\sqrt{\frac{3}{8}}b'_i \frac{V^c}{\Lambda}l_i\right) (\lambda^c_il_i) \right]
h^2_u .
\end{multline*}
Since the determinant of the inverse matrix elements vanishes, $(M^{-1}_s)_{A^\Sigma A^\Sigma} (M^{-1}_s)_{N^c N^c} - (M^{-1}_s)_{A^\Sigma N^c}^2 = (M_s)_{\bar N^c\bar N^c}/\det(M_s) = 0 $, the equation above gives again a contribution to $\theta_{13}$ but not to $\theta_{12}$ or $m_2/m_3$. The neutrino mass operator can be rewritten in fact as
\begin{equation}
\label{eq:seesaw3}
\frac{1}{4}\frac{\sigma_c}{\bar\sigma_c} \frac{1}{M_c} \left(c\lambda^c_3l'_3+\lambda^c_2 l'_2 +\frac{b'_1}{\sigma_c}\frac{M_c}{\Lambda}l'_1 \right)^2  h^2_u .
\end{equation}
In the presence of both $M_c/\Lambda$ corrections in eqs.~(\ref{eq:seesaw2},\ref{eq:seesaw3}), the total contribution to $\theta_{13}$ is
\begin{equation}
\label{eq:t13}
\theta_{13}\supset\theta^b_{13} = 2\sin\theta_{23}\frac{b_1+b'_1/(2\sigma_c)}{\lambda^c_2}\frac{M_c}{\Lambda} ,
\end{equation}
close to the experimental limit. 

% \bibliographystyle{h-physrev4}
% % \bibliographystyle{utcaps}
% \bibliography{abbrev,biblio,hep,tmp,pro}

% \end{document}

\end{document}